\documentstyle[aps,twocolumn]{revtex}
\begin{document}
\title{
Physics of the Pseudogap Phase of High $T_c$ Cuprates, or, RVB
Meets Umklapp
}
\author{Philip W. Anderson}
\address{Joseph Henry Laboratories of Physics\\
Princeton University, Princeton, N.J. 008544}
\maketitle
\begin{abstract}
It is argued that the dominant feature of the phase diagram of
the high $T_c$ cuprates is the
crossover to the pseudogap phase in the energy (temperature)
region $E (T^*)$. We argue that this
scale is determined by the effective antiferromagnetic
interaction which we calculate to be
$J_{eff} = J_{\rm superexchange} - xt$ where $x$ is the hole
percentage and $t$ the hopping
integral.
\end{abstract}

The dominant feature of the phase diagram of the high $T_c$
cuprates (see fig. 1) is the ``$T^*$
line", where the pseudogap develops.  Below $T^*$ it is generally
accepted that for low doping there
is a true antiferromagnetic region, and also generally accepted
that the superconducting ``dome"
lies below $T^*$ in the underdoped regime, but I have drawn fig.
(1) to accord to my prejudice that
the pseudogap approaches the superconducting gap in the overdoped
regime. This difference from
Nayak
et al\cite{one} and from Tallon and Loram\cite{two} is merely my
judgement of the bulk
of experimental evidence and is
not very relevant to the relatively large energy scales discussed
in this paper.  At low
temperatures and underdoping, a number of other phases may exist
in different samples; there is no
theoretical reason, for instance, why antiferromagnetism and
d-wave superconductivity cannot coexist
(see, for example, Ogata\cite{three}), the staggered-flux
(``d-density wave" state may occur, as well as
glassiness or inhomogeneity (``stripes").

But one cannot hope to understand the microscopic physics of the
competition among  these various
low-temperature fixed points until one understands the
high-temperature phase out of which they grow,
and why it occurs in this region of the phase diagram.  I take
the point of view that this phase is
an unstable fixed point (a critical point, if you like) which is
described by an effective
Hamiltonian for
the relevant degrees of freedom of charge and spin.  However, it
is still capable of further
symmetry-breaking condensations.

This is not an unfamiliar situation. We are all aware that the
Fermi liquid is such an unstable
fixed point, capable of various Cooper pair condensations
depending on the nature of the residual
(marginal) interactions. The Kondo lattice, a model for the
general mixed valence system, is an
intermediate unstable fixed point with only spin degrees of
freedom for the f electrons, which  we
know to be capable of many different condensations. This is an
example which shows how enormous
complications can arise from simple interactions.

We start from a simple 2D N-site Hubbard model. I see no reason
to believe that anything more is
essential.  There are $n=1-x$ electrons per site, and there is a
strong, local repulsive interaction
U. We keep in mind that next-neighbor hopping t' is finite, but
for simplicity, keep only t.  We
specialize to $t << U$, the physical case.
Note that near $n=0$, $x=0$, this is essentially a repulsive
free-electron model. Most believe this
renormalizes to a Fermi liquid; I don't, but the differences are
negligible for small $n$.
Thus anything interesting happens at some finite $n$, where the
lattice, i.e. the unklapp terms in the
scattering by $U$, begins to matter. Most conventional wisdom is
that only at $n=1$, $x=0$ does
anything occur, with the Mott insulator; others even deny that
and do not discuss the paramagnetic
insulator, embracing the Slater point of view that Hartree-Fock
is the only game. But I choose to
suggest that a crossover will exist where unklapp becomes
relevant, and that one view of $T^*$ is
that it is that crossover.

One way to examine the situation is via the $t-J$ Hamiltonian. At
any energy scale small compared to
$U$, it is possible to perturbatively eliminate the double
occupancy block in the Hamiltonian,
leaving one with the essentially equivalent $t-J$ Hamiltonian.
\begin{equation}
H\ =\ P \sum_{ij\sigma} t_{ij}\ c^+_{i\sigma}\ c_{j\sigma} P +
\sum_{ij} J_{ij}  S_i S_j
\ J_{ij} \simeq t^2_{ij/U}.
\end{equation}
Note that this step can be thought of as a renormalization of $U
\to \infty$: $U$ is taken as
strongly relevant and replaced by a constraint,  as in the Kondo
Hamiltonian. (P projects out
double occupancy). For $n$ near 0, $x$ near 1, the net effect is
to maintain the same two-particle
scattering amplitude for the singlet channel by adding in a
compensating attractive interaction J, but
otherwise there is no effect (for low density, the state is
entirely a function of the scattering
amplitude). But as $x \to 0$, the t-term drops out and one is
left with a pure Heisenberg
ferromagnet: a new model whose degrees of freedom are local
spins.

Let us discuss small but finite $x$ using the Hamiltonian (1).
That is, we study the doped Mott
insulator.  (1) is already on the way to being spin-charge
separated, in that only the $t$ term gives
the holes any dynamics.  The $J$ term is already of the form $F$
$(\{S_1 \cdots S_N\})$ of a
functional of the spin configuration only. Of course, there are
$x N$ missing spins, so the
S-representation is overcomplete; but in the end we will wish to
express the spin configuration in
terms of Fourier transformed variables, spinons  or if you wish
Schwinger bosons, and for the low-energy
sector the overcompleteness does not matter. The effective
Hamiltonian for the spin sector is the
trace over all initial positions of the holes.

What we need to do now is to find a functional of the spin
configuration which expresses the effect of
the $t$ term on the spins.  We recognize that we work in the
limit $t >> J$, so that hole motions are
much faster than spins. Tracing over positions, i.e. summing over
all paths for which the holes
return to the same configuration, is a kind of Born-Oppenheimer
approximation, comparable to
averaging over electron motions to obtain an effective
Hamiltonian for phonons.

The problem of obtaining the lowest order term in $x$ is almost
identical to the Nagaoka problem
discussed by Brinkman and Rice\cite{four} for the 3-dimensional
case. What is needed is the energy
distribution
of one-hole states as a functional of the state of the spins. We
need not assume, as Brinkman-Rice
did, that that state is simply expressed as a configuration of up
and down spins in a particular
direction, but leave it general, and search for an operator
functional $F$. To lowest order in $x$,
the spin state may be assumed fixed; it does not change to
accommodate a particular hole's motion,
being mostly determined by the large $J$ term. Brinkman and Rice
attacked this problem via the moment
method.  One starts from
\begin{equation}
G_{ii} (\omega) \ =\ \sum_n |(i|n)|^2 {1\over \omega - E_n} \ =\
(i | {i\over \omega - H} | i)
\end{equation}
and the latter can be expanded in terms of paths starting and
ending at site $i$ by using
\begin{equation}
{1\over \omega - H} \ =\ {1\over \omega}\ +\ {1\over \omega^2}
H\ +\ {1\over \omega^3} H^2 \ +\
\cdots
\end{equation}
It is easy to show that successive terms in (3) are the moments
of the energy distribution.

The first thing to realize is that once a spin state is chosen,
the problem contains only one energy,
$t$: the distribution has a breadth proportional to $t$,
all features are separated by energies of order $t$,
 and therefore the desired term in the energy is of  order
$tx$.  Baskaran\cite{five} attempted to derive an effective
Hamiltonian from high-temperature series, but
did
not recognize this scaling law and found terms proportional to
powers of $t$.

The unexpected discovery of Brinkman and Rice is that about 3/4
of the total width of the
distribution of states is completely independent of spin
configuration. The paths on which the hole
travels out and back along the same nonrepeating path can be
approximately summed and give a
distribution of width $2\sqrt{k}t$, whereas the width for
unobstructed notion of holes in a
ferromagnetic is $4t$. ($k$ is the ``connectivity", the
exponential rate of which nonrepeating paths
of length $N$ proliferate, about 2.5 for the square lattice.
This formula is slightly in error in
Brinkman and Rice.)

Perhaps we can think of these states made up of paths for which
the holes travel only on repeating
paths as in some sense the ``holon" band, since they are the
energies we would have for charge
fluctuations without spin fluctuations. In one dimension the
holons are exactly these states,
since all paths are Cayley tree paths in that case.
The holes actually present will primarily occupy these states at
an energy $\simeq (w^R_B)$.
This is because the spin configurations satisfying $J$ (like the
``random" and ``anti ferromagnetic"
configurations tried by Brinkman and Rice) are not friendly to
configurations with loops as we shall
see.

The rest of the hole bandwidth results from paths with loops of
the hole hoppings (see fig. 2). The
simplest of these (and the major contributor, according to the
estimates of Brinkman and Rice) is the
hopping of one hole around a single plaquet. It is easy to see
that this ``stirs" the spin
configuration and, in fact, the effect of a single tour in the
direction $4 \to 1 \to 2 \to 3 \to 4$
is proportional to
\begin{eqnarray}
P_{123} \ =\ P_{12} P_{23} &=& {1\over 4} (1 + \sigma_1 \cdot \sigma_2) (1 +
\sigma_2 \cdot \sigma_3)
{1\over 4}\nonumber \\ &+& {(\sigma_1 \cdot \sigma_2 + \sigma_2 \cdot
\sigma_3 + \sigma_3 \cdot \sigma_1) \over 4}\nonumber \\
                            &+& {1\over 4} (\sigma_1 \times \sigma_2) \cdot
\sigma_3  \end{eqnarray}
The sign of this  term and all higher terms is negative
(ferromagnetic) because all possible loops
contain an odd number of spins and hence lead to an odd
permutation, and because the energy of the
hole state is negative.

Note the term proportional to the chirality of the arrangement of
$S_1$, $S_2$, and $S_3$.
This is important in that it can become relevant when it comes
time to consider hole-hole and
hole-spin interactions, but for the time being, where we average
over all hole paths assuming the
spin configuration is fixed, it cancels because we must include
both chiralities of hole motion.
However, I suspect that this term is one of those important in
motivating the ``staggered flux"
phase which may play a role in some regions.

By rearranging the series we can insert these loop terms into the
repeating paths as a hole
self-energy.
\begin{equation}
\sum_h \ =\ - Kt \sum_{<ij>} \ S_i S_j
\end{equation}
($K$ being an unknown constant of order unity). Thus the effect
in first-order is to reduce the
effective $J$
\begin{equation}
J\ \to \ J_{superx} \ -\ Ktx
\end{equation}
The coefficient $K$ may be estimated by comparing to the pure
ferromagnetic state, where hole
motion is perfectly free and the band has width 4t. $K$ is
approximately unity.

It is of course not surprising that the $t$ term is
ferromagnetic, since it is known that the infinite
$U$ Hubbard model has a ferromagnetic ground state for $x
{<\atop\sim} .3$

(6) is the basic result of this paper.  It echoes the estimate of
$T^*$ provided by Loram   and
Tallon. We identify the pseudogap phase as being a region of
charge-spin separation in which the spin
Hamiltonian is dominated by the $J$ term, which falls to zero and
changes sign above a concentration
$x$ between .2 and .3 for the cuprates.
This estimate of the pseudogap crossover (and a diagram similar
to Fig. 1) are nicely confirmed by
recent measurements of Shibauchi et al. \cite{seven}

What is the effect of $J$? This can only be certainly described
in the high $T^*$, low $x$ regime.
For larger $x$ and low $T$ the further terms in $X^2$, and the
current-spin current couplings
such as the chiral term in (4),
begin to be important. But in the temperature region where these
are irrelevant, we know that
$J_{eff}$ acting in the spin sector alone gives us a flux-phase
or ``s+id" RVB in the spin sector,
whose excitations are nodal spinons. This comes about by pair
factorizing
\[
J \sum_{ij} \ S_i  S_j \ =\ J \sum_{\sigma\sigma'\atop ij}
s^*_{i\sigma}\ S_{\sigma\sigma'}
s_{i\sigma} s_{j\sigma'}\ S_{\sigma'\sigma} s_{j\sigma}
\]
and observing that the lowest energy such state is the s+id RVB,
equivalent to a $\pi$ flux phase.
This is an unstable fixed point at all concentrations $x$. At
$x=0$, the antiferromagnet supplants it
below $T_N$, since the spinons form bound pairs as described by
Hsu.\cite{six}  But at finite $x$, the
fluctuations due to hole motion destroy this order, as I shall
show in a forthcoming paper, and the
kinetic energy terms in the higher powers of $x$ cause
superconductivity.

\begin{figure}

\caption{My Version of the Generalized Phase Diagram}

\caption{Paths Important for the Self-Energy Sum}

\end{figure}

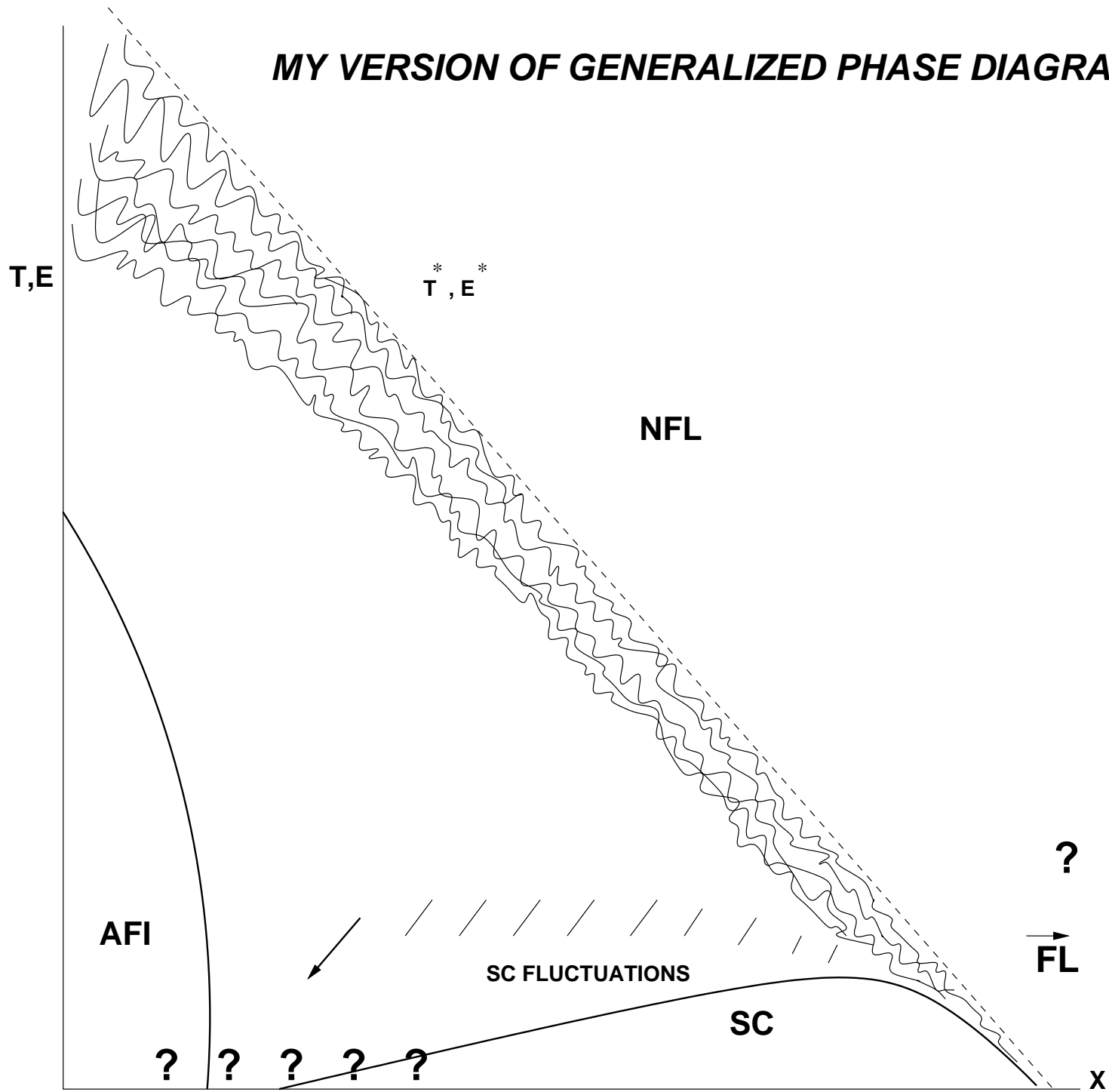
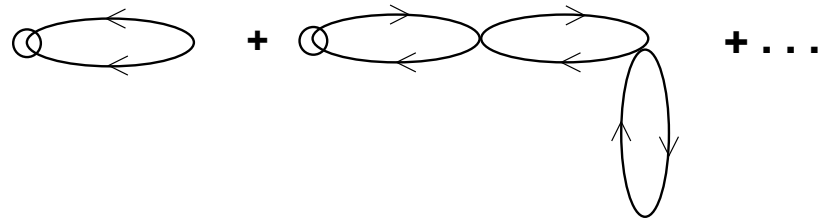
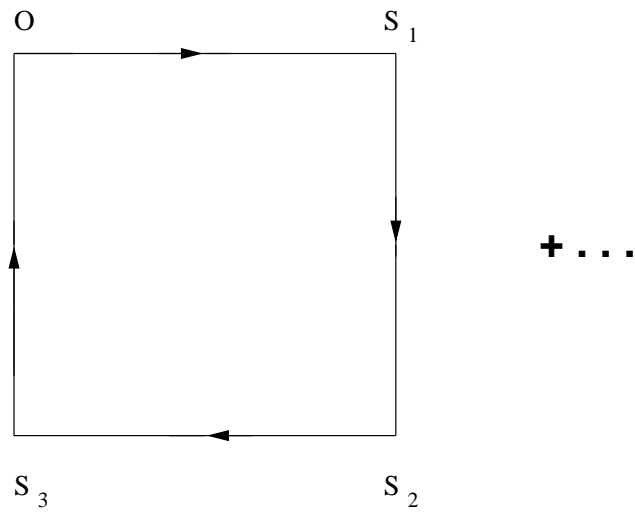
\end{document}